\documentstyle[12pt,graphicx]{article}
\begin{document}

\def\ds{\displaystyle}
\def\beq{\begin{equation}}
\def\eeq{\end{equation}}
\def\bea{\begin{eqnarray}}
\def\eea{\end{eqnarray}}
\def\beeq{\begin{eqnarray}}
\def\eeeq{\end{eqnarray}}
\def\ve{\vert}
\def\vel{\left|}
\def\ver{\right|}
\def\nnb{\nonumber}
\def\ga{\left(}
\def\dr{\right)}
\def\aga{\left\{}
\def\adr{\right\}}
\def\lla{\left<}
\def\rra{\right>}
\def\rar{\rightarrow}
\def\nnb{\nonumber}
\def\la{\langle}
\def\ra{\rangle}
\def\ba{\begin{array}}
\def\ea{\end{array}}
\def\tr{\mbox{Tr}}
\def\ssp{{\Sigma^{*+}}}
\def\sso{{\Sigma^{*0}}}
\def\ssm{{\Sigma^{*-}}}
\def\xis0{{\Xi^{*0}}}
\def\xism{{\Xi^{*-}}}
\def\qs{\la \bar s s \ra}
\def\qu{\la \bar u u \ra}
\def\qd{\la \bar d d \ra}
\def\qq{\la \bar q q \ra}
\def\gGgG{\la g^2 G^2 \ra}
\def\q{\gamma_5 \not\!q}
\def\x{\gamma_5 \not\!x}
\def\g5{\gamma_5}
\def\sb{S_Q^{cf}}
\def\sd{S_d^{be}}
\def\su{S_u^{ad}}
\def\ss{S_s^{??}}
\def\sbp{{S}_Q^{'cf}}
\def\sdp{{S}_d^{'be}}
\def\sup{{S}_u^{'ad}}
\def\ssp{{S}_s^{'??}}
\def\sig{\sigma_{\mu \nu} \gamma_5 p^\mu q^\nu}
\def\fo{f_0(\frac{s_0}{M^2})}
\def\ffi{f_1(\frac{s_0}{M^2})}
\def\fii{f_2(\frac{s_0}{M^2})}
\def\O{{\cal O}}
\def\sl{{\Sigma^0 \Lambda}}
\def\es{\!\!\! &=& \!\!\!}
\def\ar{&+& \!\!\!}
\def\ek{&-& \!\!\!}
\def\cp{&\times& \!\!\!}
\def\se{\!\!\! &\simeq& \!\!\!}
\def\kpm{&\pm& \!\!\!}
\def\kmp{&\mp& \!\!\!}


\def\simlt{\stackrel{<}{{}_\sim}}
\def\simgt{\stackrel{>}{{}_\sim}}


\title{
         {\Large
                 {\bf
Semileptonic  $B \rar \eta  \ell \nu$ decay in light cone
 QCD
                 }
         }
      }

\author{\vspace{1cm}\\
{\small T. M. Aliev$^a$ \thanks {e-mail: taliev@metu.edu.tr}\,\,,
{\. I} Kan{\i}k$^a$ \thanks {e-mail: e114288@metu.edu.tr}\,\,, A.
\"{O}zpineci$^b$ \thanks {e-mail: ozpineci@ictp.trieste.it}\,\,,}
\\
{\small a Physics Department, Middle East Technical University,
06531 Ankara, Turkey}\\
{\small b  The Abdus Salam International Center for Theoretical Physics,
I-34100, Trieste, Italy} }
\date{}

\begin{titlepage}
\maketitle
\thispagestyle{empty}

\begin{abstract}
We study semileptonic decay $B \rar \eta  \ell \nu$. The
transition form factors for this decays are calculated by using
light cone QCD sum rules. 

\end{abstract}
~~~PACS numbers:13.20 He,14.40-n
\end{titlepage}

\section{Introduction}
The paradigm of CP  violation related to the structure of CKM matrix in the Standard Model (and its beyond) is
fueling an impressive experimental programme for the study of both exclusive and inclusive
B-decays.
The data from BaBar and Belle open new era of B meson physics, and
open a real possibility for detecting semileptonic decay modes of B mesons.
The semileptonic decay modes of B mesons are much cleaner samples then the nonleptonic decay modes
since in these modes, there does not exist any  problems
connected with the presence of a third strongly interacting particle.
For this reason, the study of semileptonic  decays is  one of the efficient ways for the
determination of the CKM matrix elements.
For example $V_{cb}$ has been determined from semileptonic meson decays \cite{R1}.
CLEO collaboration \cite{R2} have measured the branching ratios of $B^0 \rar  \pi^- \ell^+ \nu$ and
$B^0 \rar \rho^- \ell^+ \nu$ which leads to  $|V_{ub}| = (3.25 \pm 0.14^{+0.21}_{-0.29} \pm 0.55) \, 10^{-3}$.
It is well known that for accurate determination of CKM matrix elements we need more reliable determination of transition form factors.

In this work we calculate the transition form factors for $B \rar
\eta  \ell \nu$ decay in light cone QCD sum rules (LCQSR) method.
The detailed description of the method and its applications can be
found in \cite{R3,R4,R5}. It should be noted that interest to
the $B \rar \eta  \ell \nu$ and $B \rar \eta'  \ell \nu$ semileptonic decays is due to the fact that they also give
information on the $\eta -\eta'$ mixing (see for example
\cite{R6,R7}).

The work is organized as follows. In Sect. 2, we derive the sum
rules for the transition form factors of $B \rar \eta \ell \nu$
decay. Section 3 is devoted to the numerical analysis and contain
a summary of the results and our conclusions.

\section{Light Cone QCD Sum Rules for the $B \rar \eta $ transition form factors}

The $B \rar \eta $ weak transition form factors $f_+(q^2)$ and
$f_-(q^2)$ are defined as: \bea \la \eta(p) | \bar u \gamma_\mu (1
+ \gamma_5) b | B(p+q) \ra = 2 f_+ p_\mu + (f_+ + f_-) q_\mu \, .
\label{eq1} \eea In this section, we calculate these form factors
using light cone QCD sum rule method. We adopt the usual strategy
of light cone sum rules method by considering the following
correlator function: \bea \Pi_\mu(p,q) = i \int d^4 x e^{i q x} \
\la \eta(p) | T\left\{ \bar u (x) \gamma_\mu (1+\gamma_5) b(x)
\bar b(0) i (1+\gamma_5) u(0) \right\} | 0 \ra \label{eq2} \eea
The main reason for choosing the chiral $\bar  b i (1+\gamma_5) u$
current instead of the $\bar b i \gamma_5 u$ current, which have
been used in the calculation of the $B \rar \pi$ weak form factor
\cite{R8}, is that in this case, twist-3 wave functions which are
the main inputs in LCQSR method and which bring the main
uncertainty to the prediction, do not contribute (see below, see
also \cite{R9}).

First let us consider the hadronic representation of the correlator. By inserting a complete set of states with the same quantum
numbers as the operator $\bar b i (1 + \gamma_5) u$ between the currents and isolating the pole term of the lowest pseudo scalar
$B$ meson, we get the following hadronic representation of the correlator:
\bea
\Pi_\mu(p,q) &=& \frac{
\la \eta | \bar u \gamma_\mu b | B \ra \la B | \bar b i \gamma_5 u | 0 \ra}{m_B^2 - (p+q)^2} +
\sum_h \frac{\la \eta | \bar u \gamma_\mu b | h \ra \la h | \bar b i (1 + \gamma_5) u | 0 \ra}{m_h^2 - (p+q)^2} \nnb \\
&=& \Pi_1(q^2, (p+q)^2) p_\mu + \Pi_2 (q^2, (p+q)^2) q_\mu
\label{eq3}
\eea

The sum in Eq. \ref{eq3} takes into account the contributions of the higher states and continuum.
Note that the intermediate states $h$ contain not only pseudo scalar resonances of masses greater the
$m_B$, but also scalar resonances with $J^P=0^+$ corresponding to the operator $\bar b u$.

For the invariant amplitudes $\Pi_i$, one can write a general
dispersion relation in the $B$ meson momentum squared, $(p+q)^2$,
as: 
\bea 
\Pi_i(q^2,(p+q)^2) = \int ds \frac{\rho_i(s)}{s-(p+q)^2} + ~subtractions
\label{eq4} 
\eea 
where $subtractions$ are polynomials in $(p+q)^2$ and the spectral densities corresponding to Eq.
(\ref{eq3}) are given as: 
\bea
\rho_1(s) &=& 2 f_+ (q^2) \frac{m_B^2 f_B}{m_b} \delta(s-m_B^2) + \rho_1^h(s) \label{eq5}\\
\rho_2(s) &=& (f_+ + f_-) \frac{m_B^2 f_B}{m_b} \delta(s-m_B^2) + \rho_2^h(s) \label{eq6}
\eea
The first terms in Eqs. (\ref{eq5}) and (\ref{eq6}) represent the ground state B-meson contribution
and are easily obtained from Eq. (\ref{eq3}) using Eq. (\ref{eq1}) and the definition
\bea
\la B | \bar b i \gamma_5 d | 0 \ra = \frac{m_B^2 f_B}{m_b} \nnb
\eea
whereas $\rho_i^h$ represent the spectral densities of the higher
resonances and the continuum. The spectral densities $\rho_i^h$ can be approximated by invoking the
quark-hadron duality ansatz:
\bea
\rho_i^h(s) = \rho_i^{QCD} \theta(s-s_0)
\label{eq7}
\eea
So for the hadronic representation of the invariant amplitudes $\Pi_i$ we get:
\bea
\Pi_1 &=& \frac{2 f_+(q^2) m_B^2 f_B}{m_b (m_B^2 - (p+q)^2)} + \int_{s_0}^\infty ds
\frac{\rho_1^{QCD}(s)}{s-(p+q)^2} + \, subtractions \nnb \\
\Pi_2 &=& \frac{(f_+ + f_-) m_B^2 f_B}{m_b (m_B^2 - (p+q)^2)} + \int_{s_0}^\infty ds
\frac{\rho_2^{QCD}(s)}{s-(p+q)^2} + \, subtractions
\label{eq8}
\eea

For obtaining sum rules for the form factors $f_+$ and $f_-$ we
must calculate the correlator function in QCD. It can be done by
using the light cone OPE method; i.e. by expanding the $T$ product
of currents near the light cone $x^2 \simeq 0$. After contracting
the $b$-quark fields we get \bea \Pi_\mu(p,q) = i \int d^4 x e^{i
q x} \la \eta(p) | \bar u(x) \gamma_\mu (1 + \gamma_5) S_b(x) (1 +
\gamma_5) u(0) | 0 \ra \label{eq9} \eea where $S_b$ is the full
propagator of the $b$ quark. Its explicit expression is given by
the following expression \bea i S_b(x) &=& i S_b^0(x) - i gs \int
\frac{d^4 k}{(2 \pi)^4} e^{-i k x} \int_0^2 dv \left\{ \frac{1}{2}
\frac{\not\!k + m_b}{(m_b^2 - k^2)^2} G^{\mu \nu}(v x) \sigma_{\mu
\nu} \right. \nnb \\ && + \left. \frac{1}{m_b^2 - k^2} v x_\mu
G^{\mu \nu}(v x) \gamma_\nu \right\} \label{eq10} \eea Here
$S_b^0(x)$ is the free quark propagator of the massive $b$ quark:
\bea S_b^0(x) = \frac{m_b^2}{4 \pi^2} \frac{K_1(m_b
\sqrt{-x^2})}{\sqrt{-x^2}} - i \frac{m_b^2}{4 \pi^2}
\frac{\not\!x}{x^2} K_2(m_b \sqrt{-x^2}) \label{eq11} \eea where
$K_i$ are the Bessel functions.

From Eqs. (\ref{eq9})-(\ref{eq11}) it follows that, in order to calculate the theoretical part of
the correlator function, the matrix elements of the nonlocal operators between $\eta$ meson and
vacuum states are needed.

Here we would like to do following remark: in all the following calculations
we neglect the $\eta-\eta'$ mixing since it is very small \cite{R1}, and therefore we choose
as the interpolating current for the $\eta$ meson, the  SU(3) octet axial vector
current: 
\bea 
J_\mu =
\frac{1}{\sqrt{6}}(\overline{u}\gamma_{\mu}\gamma_{5}u+\overline{d}\gamma_{\mu}\gamma_{5}d-2\overline{s}\gamma_{\mu}\gamma_{5}s)\eea

In order to simplify the notation we use
$\overline{q}\Gamma_{\mu}q$ to denote
$\frac{1}{\sqrt{6}}(\overline{u}\gamma_{\mu}\gamma_{5}u+\overline{d}\gamma_{\mu}\gamma_{5}d-2\overline{s}\gamma_{\mu}\gamma_{5}s)$
and introduce notation $F_\eta = \frac{f_\eta}{\sqrt{6}}$, where
$f_\eta$ determined as $\la 0 | \overline{q}(0)\Gamma_{\mu} q(0) |
\eta(q) \ra = -if_\eta q_\mu$.

 An important observation is that the terms in $S_b(x)$
containing odd number of gamma matrices do not contribute, i.e.
leading twist-3 terms do not give any contribution. Up to twist-4,
the $\eta$ meson wave functions are defined in  the following way
\cite{R10} : \bea \la \eta(p) | \bar q \gamma_\mu \gamma_5 q | 0
\ra &=& -i f_\eta p_\mu \int_0^1 du e^{-i u p x} \left[
\varphi_\eta (u) + \frac{1}{16} m_\eta^2 x^2 A(u) \right] \nnb \\
&& - \frac{i}{2} f_\eta m_\eta^2 \frac{x_\mu}{px} \int_0^1 du
e^{-i u px} B(u)
\label{eq12}\\
\la \eta(p) | \bar q(x) \gamma_\mu \gamma_5 g_s G_{\alpha \beta}(v x) q(0) | 0 \ra &=&
f_\eta m_\eta^2 \left[ p_\beta \left( g_{\alpha \mu} - \frac{x_\alpha p_\mu}{px} \right)
-p_\alpha \left( g_{\beta \mu} - \frac{x_\beta p_\mu}{px} \right) \right]
\nnb \\ &&
\times \int {\cal D}\alpha_i \varphi_\perp(\alpha_i) e^{-i p x(\alpha_1 + u \alpha_3)}
\nnb \\ &&
+ f_\eta m_\eta^2 \frac{p_\mu}{px} (p_\alpha x_\beta - p_\beta x_\alpha)
\int {\cal D}\alpha_i \varphi_\parallel(\alpha_i) e^{-i p x(\alpha_1 + u \alpha_3)}
\nnb \\
\label{eq13}\\
\la \eta(p) | \bar q(x) g_s \tilde G_{\alpha \beta}(vx) \gamma_\mu q(0) | 0 \ra &=&
i f_\eta m_\eta^2 \left[ p_\beta \left( g_{\alpha \mu} - \frac{x_\alpha p_\mu}{px} \right)
-p_\alpha \left( g_{\beta \mu} - \frac{x_\beta p_\mu}{px} \right) \right]
\nnb \\ &&
\times \int {\cal D}\alpha_i \tilde \varphi_\perp(\alpha_i) e^{-i p x(\alpha_1 + u \alpha_3)}
\nnb \\ &&
+i f_\eta m_\eta^2 \frac{p_\mu}{px} (p_\alpha x_\beta - p_\beta x_\alpha)
\int {\cal D}\alpha_i \tilde \varphi_\parallel(\alpha_i) e^{-i p x(\alpha_1 + u \alpha_3)}
\nnb \\
\label{eq14}
\eea
where $\tilde G_{\mu \nu} = \frac{1}{2} \epsilon_{\mu \nu \sigma \lambda} G^{\sigma \lambda}$, ${\cal
D} \alpha_i = d \alpha_1 d \alpha_2 d \alpha_3 \delta(1-\alpha_1 - \alpha_2 - \alpha_3)$.
In Eqs. (\ref{eq12})-(\ref{eq14}), the $\varphi_\eta(u)$ is the leading twist-2, $A(u)$ and part of
$B(u)$ are two particle twist-4, $\varphi_\parallel(\alpha_i)$, $\varphi_\perp(\alpha_i)$, $\tilde
\varphi_\parallel(\alpha_i)$ and $\tilde \varphi_\perp(\alpha_i)$ are three particle twist-4 wave
functions.

Here we should note that that matrix element
$\la \eta(p) | \bar u \gamma_\mu G^{\alpha \beta}(u x) \sigma_{\alpha \beta} u | 0 \ra = 0$
due to the parity invariance of strong interactions and using the identity
\bea
\gamma_\mu \sigma_{\alpha \beta} = i (g_{\mu \alpha} \gamma_\beta - g_{\mu \beta} \gamma_\alpha ) +
\epsilon_{\mu \alpha \beta \rho} \gamma^\rho \gamma_5
\nnb
\eea

Inserting Eqs. (\ref{eq10}) and (\ref{eq11}) into Eq. (\ref{eq9})
and using definitions of $\eta$-meson wave functions (Eqs.
(\ref{eq12})-(\ref{eq14})) for the theoretical part of the
correlation function we get: 
\bea 
\Pi^{th}_\mu = && i \sqrt2 \int d^4 x
e^{i q x} \left[ \int_0^1 du e^{i u p x} \left\{ -\frac{1}{32
\pi^2} F_\eta m_b^2 (16 \varphi_\eta + A m_\eta^2 x^2)
\frac{K_1(m_b \sqrt{-x^2})}{\sqrt{-x^2}} p_\mu\right. \right. \nnb
\\ && \left. - \frac{1}{4 \pi^2} F_\eta m_b^2 m_\eta^2 B
\frac{K_1(m_b \sqrt{-x^2})}{\sqrt{-x^2}} \frac{x_\mu}{p
 x} \right\}
\nnb \\ && + \int {\cal D} \alpha \int_0^1 du e^{i (\alpha_2 + u
\alpha_3) px} \left\{ \frac{1}{12 \pi^2} F_\eta m_\eta^2 m_b
(\varphi_\parallel + \tilde \varphi_\parallel)  K_0(m_b
\sqrt{-x^2}) \frac{ x_\mu p^2 - p_\mu (p  x) }{p  x} \right. \nnb
\\ && \left. \left. + \frac{1}{12 \pi^2} F_\eta m_\eta^2 m_b
(\varphi_\perp + \tilde \varphi_\perp) K_0(m_b \sqrt{-x^2})
\frac{p^2 x_\mu + 2 (p  x) p_\mu}{p  x} \right\} \right] \eea The
next task is to carry out the fourier transformation and then to
take the Borel transform with respect to the variable $(p+q)^2$ in
order to suppress the contributions of the higher states and the
continuum and also in order to eliminate the subtraction terms.
Finally we identify the same structures both at the hadronic and
quark gluon levels. Subtraction of the continuum contribution is
done by employing the quark-hadron duality as mentioned before.
This amounts to restricting the spectral integral to $s \le s_0$.
(for more details about this procedure see the appendix and also
\cite{R8}). For the theoretical part of the correlator
(\ref{eq2}) function we get: \bea \Pi^{th}_\mu &=& \sqrt2 F_\eta
\frac{m_\eta^2}{3} I_1^1(\varphi_\parallel + \tilde
\varphi_\parallel - 2 \varphi_\perp - 2 \tilde \varphi_\perp)
p_\mu +  \sqrt2 F_\eta \frac{m_\eta^4}{3 m_b} \tilde
I_2^2(\varphi_\parallel + \tilde \varphi_\parallel+\varphi_\perp
+\tilde \varphi_\perp) p_\mu \nnb \\ && + 2 \sqrt2 F_\eta m_b
J_1^0(\varphi_\eta) p_\mu - \sqrt2 \frac{F_\eta}{16} m_b  m_\eta^2
J_1^2(A) p_\mu - \sqrt2 F_\eta m_\eta^2 \tilde J_2^1(B) p_\mu \nnb \\ && -
\sqrt2 F_\eta m_\eta^2 J_2^1(B) q_\mu + \frac{\sqrt2}{3 m_b} F_\eta
m_\eta^4  I_2^2(\varphi_\parallel + \tilde
\varphi_\parallel+\varphi_\perp +\tilde \varphi_\perp) q_\mu
\label{eq16}
\eea
where the functions $I_n^m(\varphi)$, $J_n^m(\varphi)$, $\tilde I_n^m(\varphi)$, and $\tilde J_n^m(\varphi)$
are defined in the appendix.

Equating the coefficients of the corresponding $p_\mu$ and $q_\mu$ structures in Eqs.
(\ref{eq3}-\ref{eq6}))
and (\ref{eq16}) we get the following sum rules for the form factors $f_+$ and $f_+ + f_-$
respectively:
\bea
f_+ &=& \frac{m_b F_\eta}{\sqrt 2 m_B^2 f_B} e^{\frac{m_B^2}{M^2}} \left[
\frac{m_\eta^2}{3} I_1^1(\varphi_\parallel + \tilde \varphi_\parallel - 2 \varphi_\perp - 2 \tilde \varphi_\perp)
\right. \nnb \\  && \left.
+ \frac{m_\eta^4}{3 m_b} \tilde I_2^2(\varphi_\parallel + \tilde \varphi_\parallel+\varphi_\perp +\tilde \varphi_\perp)
+ 2 m_b J_1^0(\varphi_\eta)
- \frac{m_b m_\eta^2}{16}J_1^2(A)
- m_\eta^2  \tilde J_2^1(B) \right]
\label{eq17} \\
f_+ + f_- &=& \frac{\sqrt2 F_\eta m_b}{m_B^2 f_B} e^{\frac{m_B^2}{M^2}} \left[
- m_\eta^2J_2^1(B)
+ \frac{m_\eta^4}{3 m_b} I_2^2(\varphi_\parallel + \tilde \varphi_\parallel+\varphi_\perp +\tilde \varphi_\perp)
\right]
\label{eq18}
\eea
Eqs. (\ref{eq17}) and (\ref{eq18}) are our final results for the $B \rar \eta$ transition
form factors.

\section{Numerical Analysis}

From Eqs. (\ref{eq17}) and (\ref{eq18}) it follows that the main input parameters of these sum rules
are the $\eta$ meson wave functions. The explicit expressions of the wave functions
$\varphi_\eta(u)$, $A(u)$, $B(u)$ and $\varphi_\parallel(\alpha_i)$, $\varphi_\perp(\alpha_i)$,
$\tilde \varphi_\parallel(\alpha_i)$, and $\tilde \varphi_\perp(\alpha_i)$ are given in \cite{R10}.

The next input parameter of the sum rule is leptonic decay
constant $F_\eta$. It should be noted that for $\eta$ meson case,
the situation is more complicated compared to the $\pi$ and $K$
meson cases, due to the mixing with $\eta'$. As we already noted
that we neglect mixing between $\eta$ and $\eta'$. More recent
analysis show that the coupling to the octet current $\eta$ is
$f_\eta = 159~MeV$ \cite{R20}which we use in our numerical
calculations. 

In our calculations we neglect QCD radiative corrections.
Therefore for consistency we neglect QCD corrections to the
leptonic decay $f_B=160~MeV$ \cite{R8} (see also \cite{R17}).

Having these input parameters, we can carry out our numerical
calculation. According to QCD sum rules philosophy, first of all,
a suitable range for the auxiliary Borel parameter $M^2$ should be
found such that the numerical results are stable. The lower limit
of $M^2$ is determined by the requirement that the terms $\sim
M^{-2 n}\, ,(n>1)$ remain sub dominant. The upper bound of $M^2$ is
determined by requiring that the contributions of the higher
resonances and the continuum are less then $30\%$ of the total
result. Our numerical calculations leads that both requirements
are satisfied in the region $10\, GeV^2 < M^2 < 16 \, GeV^2$. 
It should be noted that the
light cone QCD sum rules predictions for the form factors are
reliable at the region 
\bea 
q^2 \simlt m_b^2 - 2 m_b \chi \nnb 
\eea
where $\chi $ is a typical hadronic scale of roughly $0.5\, GeV$.
Then $q^2 \simlt 18 \, GeV^2$.

\begin{figure}[h!]
$\left. \right.$
\vspace{-1cm}
\begin{center}
\includegraphics[width=10cm,angle=-90]{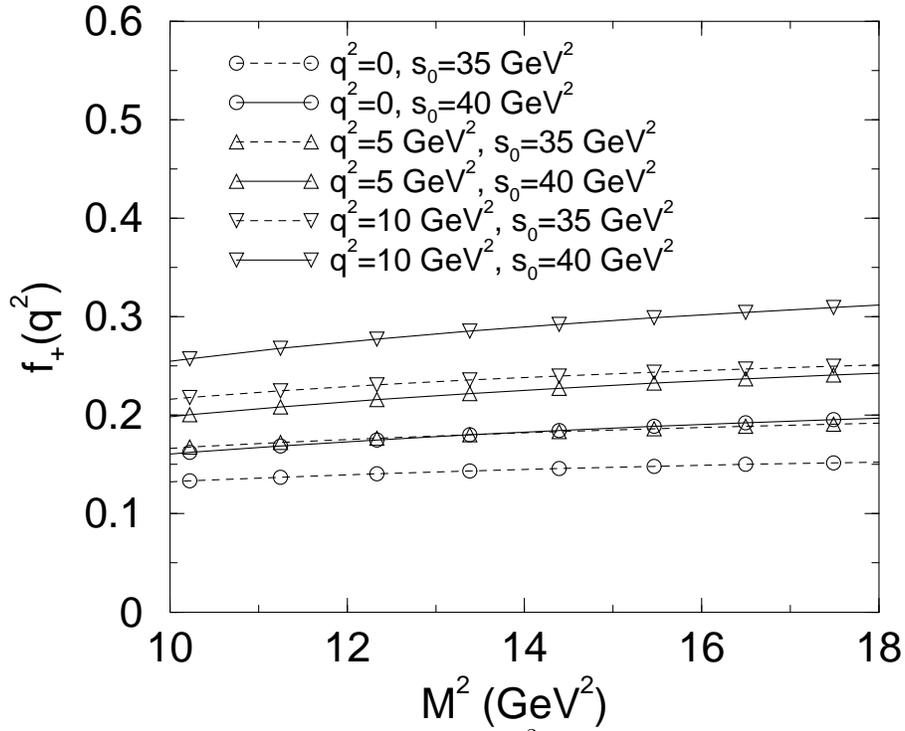}
\end{center}
\vspace{-1cm}
\caption{The dependence of the form factor $f_+(q^2)$ on the Borel parameter $M^2$ at $q^2=0$, $5$,
and $10~GeV^2$ for $s_0=35~GeV^2$ and $s_0=40~GeV^2$.}
\label{fp.msq}
\end{figure}

\begin{figure}[h!]
$\left. \right.$
\vspace{-1cm}
\begin{center}
\includegraphics[width=10cm,angle=-90]{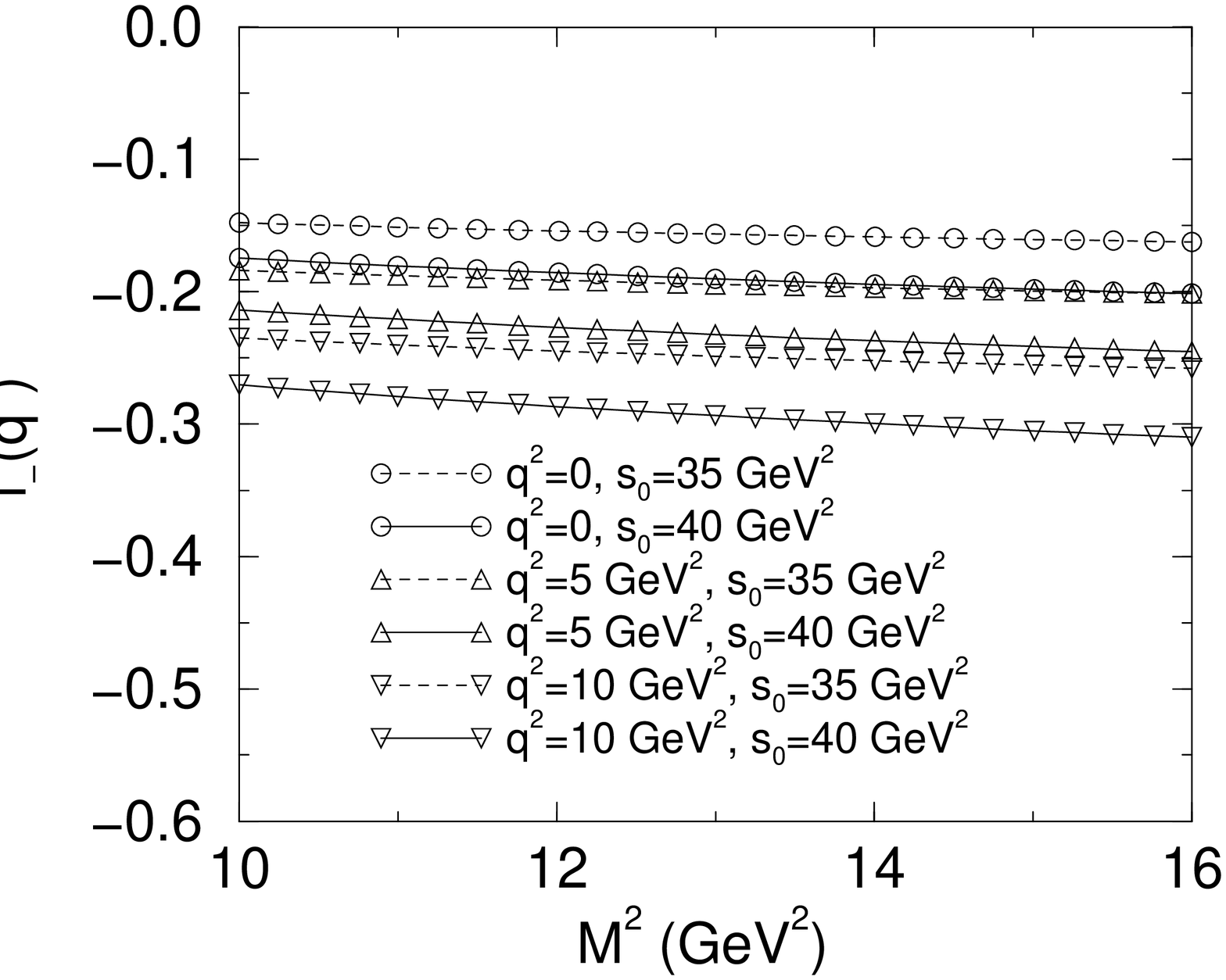}
\end{center}
\vspace{-1cm}
\caption{The same as Fig. (\ref{fp.msq}) but for the form factor $f_-(q^2)$.}
\label{fm.msq}
\end{figure}

In Figs. (\ref{fp.msq}) and (\ref{fm.msq}), the $M^2$ dependencies of the form factors $f_+(q^2)$ and
$f_-(q^2)$ is depicted for three different values of $q^2=0\, ,~5\,, ~10~GeV^2$ for two different
values of the continuum threshold $s_0=35\, , ~40~GeV^2$. From these figures, we see that $f_+(q^2)$
and $f_-(q^2)$ vary weakly as $M^2$ varies in the region $10\, GeV^2 < M^2
< 16 \, GeV^2$ up to $q^2 = 18 \, GeV^2$.

Having the working region of $M^2$, our next problem is to find
the dependence of the form factors $f_+(f_-)(q^2)$ on $q^2$ at
given values of $M^2$, namely at $M^2=12\, GeV^2$ and $M^2 = 16\,
GeV^2$, at $s_0=35\, GeV^2$ and $s_0=40 \, GeV^2$. We find that
$f_+^\eta(0) = 0.15 \pm 0.3$ and $f_-^\eta(0) = -0.16 \pm 0.2$. As
we already noted, in  the region $q^2 \ge 18 \, GeV^2$,
applicability of the light cone QCD sum rules is questionable. In
order to extend our results to the full physical region, we look
some parameterization of the form factors in such a way that in the
region, $0 \le q^2 \le 18 \, GeV^2$, this parameterization coincide
with light cone QCD sum rules predictions. A good parameterization
of the $q^2$ dependence can be given in terms of three parameters
as: \bea f_i(q^2) = \frac{f_i(0)}{1 -
a_F\left(\frac{q^2}{m_B^2}\right) + b_F
\left(\frac{q^2}{m_B^2}\right)^2} \eea The values of these
parameters for the $B \rar \eta$  transition form factors $f_+ (f_-
)$ are obtained as $a_F = 1.07 \pm 0.08$ ($a_F = 1.1 \pm 0.1$) and $b_F =
0.19 \pm 0.16$ ($b_F = 0.28 \pm 0.19$) where the quoted errors are only due to
the variations of the continuum threshold, $s_0$, and the Borel
mass, $M^2$.

It should be noted that the form factors of $B \rar \eta$
transition can be related from $B \rar \pi$ transition form factors
using $SU(3)_F$ symmetry. For example, the value of $f_+ (q^2 =0) = 0.15$
 obtained via $SU(3)_F$ symmetry is consistent with our prediction
of $f_+ (q^2 =0)$.

Finally, we would like to note that the background for the $B \rar
\eta\ell \nu$ decay would be much smaller than that for the $B
\rar \pi \ell \nu$ decay, due to the much lower multiplicity,
since the background caused by $B \rar \eta X$ is one order of
magnitude smaller than that of $B \rar \pi X$.

In conclusion, we calculate the transition form factors for $B \rar
\eta \ell \nu$ decay using the light cone QCD sum rules.

\appendix
\setcounter{equation}{0}
\renewcommand{\theequation}{A\arabic{equation}}
\section{Appendix}
In this appendix, the derivations of the  explicit forms of the functions,
$I_i$, and $J_i$, appearing in Eq. (\ref{eq16}) are presented and
the method to calculate the continuum subtractions is
demonstrated. For this purpose, consider a general term of the
correlation function : 
\bea 
\Pi(p,q) = \int d^4x e^{i qx}\int_0^1
du \int{\cal D} \alpha_i e^{i(\alpha_2 + u \alpha_3)px}
\varphi(\alpha_i) f(u) \frac{K_\nu(m_b
\sqrt{-x^2})}{(\sqrt{-x^2})^\nu} \label{a1} 
\eea 
where ${\cal D}
\alpha_i = d \alpha_1 d \alpha_2 d \alpha_3 \delta(1-\alpha_1 -
\alpha_2 - \alpha_3)$ After switching to Euclidean space and
carrying out the $x$ integration, one obtains: 
\bea \Pi(p,q) = -i
\frac{2 \pi^2}{m_b^2} \int du \int {\cal D}\alpha_i dt t^{1-\nu}
\varphi(\alpha_i) f(u) e^{-\frac{t}{2 m_b}(Q^2 + m_b^2)}
\label{a2} 
\eea 
where $Q^2 = (q + k p)^2 = q^2 \bar k - p^2 k \bar
k + k (q+p)^2$ and $k = \alpha_2 + u \alpha_3$, $\bar k = 1-k$,
and all the appearing momenta are Euclidean. In order to obtain
Eq. (\ref{a2}), the following representation of the Bessel
function $K_\nu(x)$ is used: \bea \frac{K_\nu(m_b
\sqrt{x_E^2})}{(\sqrt{x_E^2})^\nu} = \frac12 \int_0^\infty
\frac{dt}{t^{\nu + 1}} \exp\left[-\frac{m_b}{2}\left(t +
\frac{x_E^2}{t}\right)\right] \label{a3} \eea where $x_E$ is the
Euclidean position vector. Carrying out the Borel transformation
with respect to $(p+q)^2$ using the identity: \bea {\cal B}e^{-
\alpha (p+q)^2} = \delta\left(\frac{1}{M^2} - \alpha\right)
\label{a4} \eea where ${\cal B}$ stands for the Borel
transformation and $M^2$ is the Borel parameter and carrying out
the $t$ integral, one obtains: \bea \Pi(M^2) = - \frac{4 i
\pi^2}{m_b} \int du \int {\cal D} \alpha_i \left( \frac{2
m_b}{kM^2} \right)^{1-\nu} e^{-s(k)} \frac{\varphi(\alpha_i)
f(u)}{k} \label{a5} \eea where \bea s(k)  = \frac{m_b^2 - q^2 \bar
k + p^2 k \bar k}{k} \label{a6} \eea where we have switched back
to Minkowskian spacetime. Note that one can write the factors
$\frac{2 m_b}{k M^2}$ is the integrand as a differential operator
acting on the exponential in the integrand and hence: \bea
\Pi(M^2) = - \frac{4 i \pi^2}{m_b} \left( -
\frac{\partial}{\partial m_b} \right)^{1-\nu} \int du {\cal D}
\alpha_i e^{-s(k)} \frac{\varphi(\alpha_i) f(u)}{k} \label{a7}
\eea

Note that
\bea
e^{- \frac{s(k)}{M^2}} = \int_0^\infty ds e^{-\frac{s}{M^2}} \delta(s- s(k))
\label{a8}
\eea
which is nothing but the spectral representation of the exponential. The contributions of the higher
states and the continuum are subtracted by replacing a cutoff, $s_0$,  instead of the infinity as the
upper limit of the spectral integral. Hence the effect of subtraction of the continuum and the
higher states is to restrict the integration region in Eq. (\ref{a7}) to the regions of $k$ for
which $0 \le s(k) \le s_0$ which implies that $\delta \le k \le 1$ where
\bea
\delta = \frac{m_\eta^2 + q^2 - s_0 + \sqrt{(m_\eta^2 + q^2 - s_0)^2 - 4 m_\eta^2 (q^2 - m_b^2)}}{2 m_\eta^2}
\label{a9}
\eea
where $p^2 = m_\eta^2$.
Hence the contributions of the term given in Eq. (\ref{a1}) to the sum rules after Borel
transformation and the continuum is subtracted is given by:
\bea
\Pi(M^2) = - \frac{4 i \pi^2}{m_b} \left( - \frac{\partial}{\partial m_b} \right)^{1-\nu}
\int du {\cal D}\alpha_i \frac{\varphi(\alpha_i) f(u)}{k} e^{- \frac{s(k)}{M^2}}
\label{a10}
\eea
where the integration region is determined by the conditions:
\bea
\delta \le k \le 1 \nnb \\
0 \le u\, , ~\alpha_i \le 1 \label{a11}
\eea
and the Dirac's delta function in the definition of ${\cal D} \alpha_i$ fixes $\sum_i \alpha_i = 1$.

In order to obtain the contributions to the sum rules from terms
containing additional powers of $x_\mu$ one can use the trick of
replacing $x_\mu$ by differentiation with respect to $q_\mu$. For
terms containing a factor of $px$ in the denominator, one can use
the following trick: in order not to have any singularity at $px =
0$, the integral of these wave functions in the absence of the
exponential should cancel. Hence, for these terms only, one
can write: 
\bea e^{i \alpha px} \rightarrow e^{i \alpha px} - 1 =
i px \int_0^\alpha dk e^{i k px} 
\eea and the rest of the
calculation is similar to the presented one. Note that the
subtracted $1$ does not contribute.

When the contributions of the continuum is subtracted and the Borel transformation is carried out
for the other terms, the following functions which appear in Eq. (\ref{eq16}) are encountered:
\bea
I_1^n(\varphi) &=& \left(- \frac{\partial}{\partial m_b} \right)^n \left[
\int_0^{1-\delta} d \alpha_1 \int_0^{1-\delta -\alpha_1} d \alpha_3 \int_0^1 du
 + \int_0^{1-\delta} d \alpha_1 \int_{1-\delta - \alpha_1}^{1-\alpha_1} \int_\frac{\delta - 1 +
\alpha_1 + \alpha_3}{\alpha_3}^1 du \right]
\nnb \\ &&
\frac{\varphi(\alpha_1,1-\alpha_1-\alpha_3,\alpha_3)}{1-\alpha_1 -\alpha_3 + u \alpha_3}
e^{- \frac{s(1-\alpha_1-\alpha_3 + u \alpha_3)}{M^2}}
\\
I_2^n(\varphi) &=& \left(- \frac{\partial}{\partial m_b} \right)^n
 \left[
\int_0^{1-\delta} d \alpha_1 \int_0^{1-\delta-\alpha_1} d \alpha_3 \int_0^1 du
\int_\delta^{1-\alpha1 - \alpha_3 + u \alpha_3} dk
\right. \nnb \\ && \left.
+  \int_0^{1-\delta} d \alpha_1 \int_{1-\delta - \alpha_1}^{1-\alpha_1} d \alpha_3
\int_\frac{\delta -1 + \alpha_1 + \alpha_3}{\alpha_3}^1 du \int_\delta^{1-\alpha_1 - \alpha_3 + u
\alpha_3} dk \right]
\nnb \\ &&
\frac{\varphi(\alpha_1,1-\alpha_1-\alpha_3,\alpha_3)}{1-\alpha_1 -\alpha_3 + u \alpha_3}
e^{- \frac{s(1-\alpha_1-\alpha_3 + u \alpha_3)}{M^2}}
\\
\tilde I_2^n(\varphi) &=& \left(- \frac{\partial}{\partial m_b} \right)^n
\left[
\int_0^{1-\delta} d \alpha_1 \int_0^{1-\delta-\alpha_1} d \alpha_3 \int_0^1 du
\int_\delta^{1-\alpha1 - \alpha_3 + u \alpha_3} dk
\right. \nnb \\ &&\left.
 + \int_0^{1-\delta} d \alpha_1 \int_{1-\delta - \alpha_1}^{1-\alpha_1} d \alpha_3
\int_\frac{\delta -1 + \alpha_1 + \alpha_3}{\alpha_3}^1 du \int_\delta^{1-\alpha_1 - \alpha_3 + u
\alpha_3} dk \right]
\nnb \\ &&\
\varphi(\alpha_1,1-\alpha_1-\alpha_3,\alpha_3)
e^{- \frac{s(1-\alpha_1-\alpha_3 + u \alpha_3)}{M^2}}
\\
J_1^n(\varphi) &=& \left(- \frac{\partial}{\partial m_b} \right)^n
\int_\delta^1 du \frac{\varphi(u)}{u} e^{-\frac{s(u)}{M^2}}
\\
J_2^n(\varphi) &=& \left(- \frac{\partial}{\partial m_b} \right)^n
\int_\delta^1 du \int_\delta^u dk \frac{\varphi(u)}{k} e^{-\frac{s(k)}{M^2}}
\\
\tilde J_2^n(\varphi) &=& \left(- \frac{\partial}{\partial m_b} \right)^n
\int_\delta^1 du \int_\delta^u dk \varphi(u) e^{-\frac{s(k)}{M^2}}
\eea

\newpage

\begin{figure}[h!]
\begin{center}
\includegraphics[width=10cm,angle=-90]{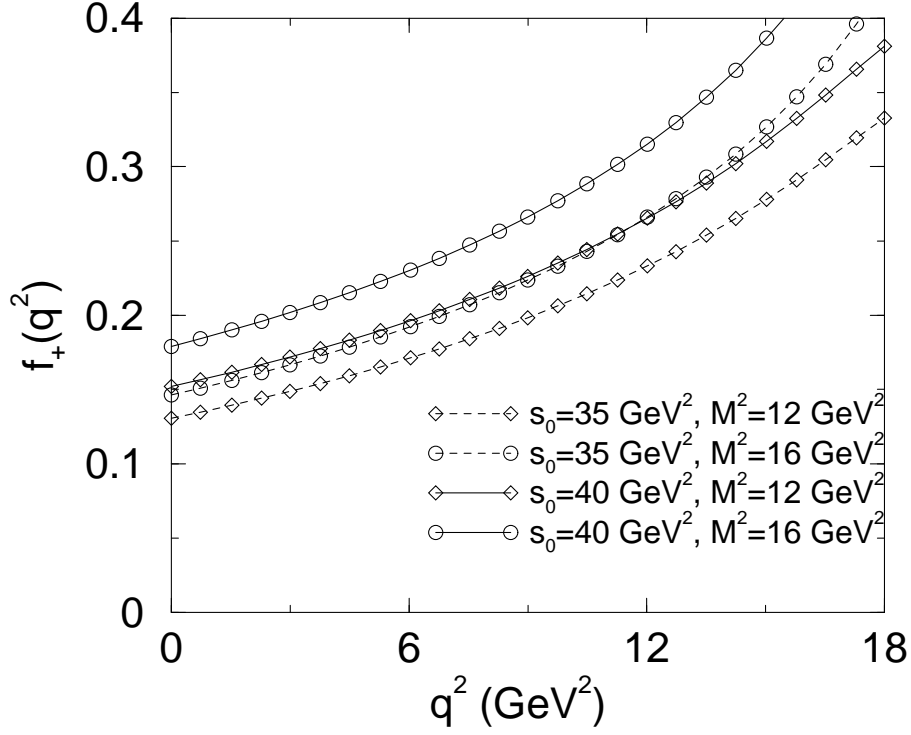}
\end{center}
\vspace{-1cm}
\caption{The dependence of the form factor $f_+(q^2)$ on $q^2$ at the continuum threshold $s_0=35~GeV^2$ and $s_0=40~GeV^2$
and at the Borel mass $M^2=12~GeV^2$ and $M^2=16~GeV^2$.}
\label{fp.qsq}
\end{figure}

\begin{figure}[h!]
$\left. \right.$
\vspace{-1cm}
\begin{center}
\includegraphics[width=10cm,angle=-90]{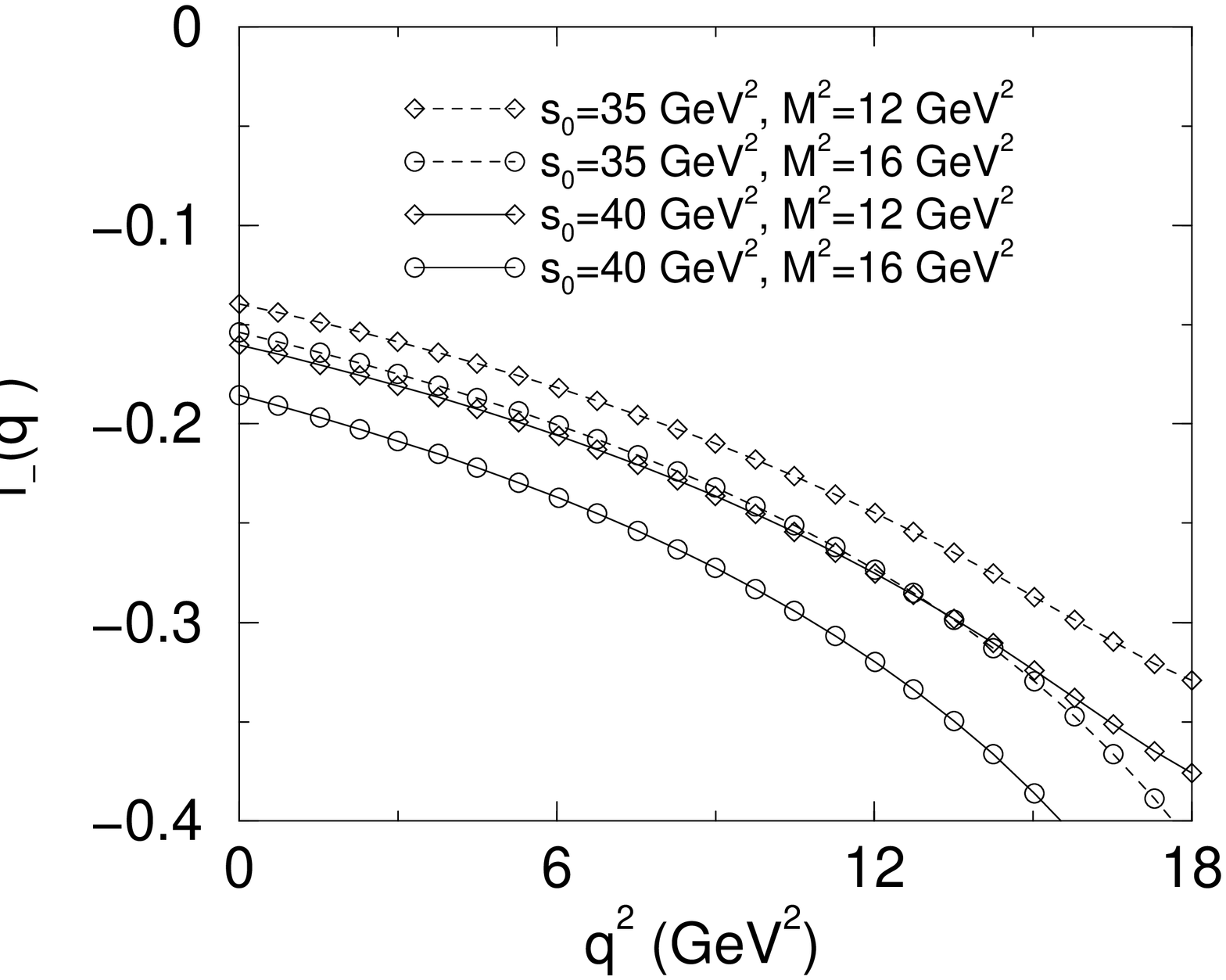}
\end{center}
\vspace{-1cm}
\caption{The same as Fig. (\ref{fp.qsq}) but for the form factor $f_-(q^2)$.}
\label{fm.qsq}
\end{figure}

\end{document}